\documentclass[sigconf,nonacm , dvipsnames]{acmart}
\usepackage{booktabs}
\usepackage{multirow}
\usepackage{pifont}
\usepackage[table]{xcolor}
\usepackage{tcolorbox}
\tcbuselibrary{breakable}
\usepackage{booktabs}
\usepackage{placeins}
\usepackage{colortbl}

\usepackage{mdframed}
\usepackage{xcolor}

\definecolor{takeawayBg}{RGB}{240,240,240}
\definecolor{takeawayRule}{RGB}{100,100,100}

\mdfdefinestyle{takeawaystyle}{
  backgroundcolor=gray!12,
  linecolor=gray!50,
  linewidth=0.8pt,
  roundcorner=5pt,
  innerleftmargin=8pt,
  innerrightmargin=8pt,
  innertopmargin=7pt,
  innerbottommargin=7pt,
  skipabove=6pt,
  skipbelow=6pt,
}

\mdfdefinestyle{llmprompt}{
  backgroundcolor=gray!5,
  linecolor=gray!40,
  linewidth=0.4pt,
  innerleftmargin=6pt,
  innerrightmargin=6pt,
  innertopmargin=4pt,
  innerbottommargin=4pt,
  frametitlebackgroundcolor=gray!15,
  frametitleaboveskip=3pt,
  frametitlebelowskip=3pt,
  frametitlerule=true,
  frametitlerulecolor=gray!40,
  frametitlerulewidth=0.4pt,
  font=\scriptsize,
}

\AtBeginDocument{%
  }

\begin{document}

\title{Defective Task Descriptions in LLM-Based Code Generation: Detection and Analysis}


\author{Amal Akli}
\orcid{}
\affiliation{%
  \institution{University of Luxembourg}
  \city{}
  \country{Luxembourg}}
\email{amal.akli@uni.lu} 

\author{Mike Papadakis}
\affiliation{%
  \institution{University of Luxembourg}
  \city{}
  \country{Luxembourg}}
\email{michail.papadakis@uni.lu}

\author{Maxime Cordy}
\orcid{}
\affiliation{%
  \institution{University of Luxembourg}
  \city{}
  \country{Luxembourg}}
\email{maxime.cordy@uni.lu}

\author{Yves Le Traon}
\affiliation{%
  \institution{University of Luxembourg}
  \city{}
  \country{Luxembourg}}
\email{Yves.LeTraon@uni.lu}
\renewcommand{\shortauthors}{Akli et al.}

\begin{abstract}
Large language models are widely used for code generation, yet they rely on an implicit assumption that the task descriptions are sufficiently detailed and well-formed. However, in practice, users may provide defective descriptions, which can have a strong effect on code correctness. To address this issue, we develop \textit{SpecValidator}, a lightweight classifier based on a small model that has been parameter-efficiently finetuned, to automatically detect task description defects. We evaluate SpecValidator on three types of defects, Lexical Vagueness, Under-Specification and Syntax-Formatting on 3 benchmarks with task descriptions of varying structure and complexity. Our results show that SpecValidator achieves defect detection of F1 = 0.804 and MCC = 0.745, significantly outperforming GPT-5-mini (F1 = 0.469 and MCC = 0.281) and Claude Sonnet~4 (F1 = 0.518 and MCC = 0.359). Perhaps more importantly, our analysis indicates that SpecValidator can generalize to unseen issues and detect unknown Under-Specification defects in the original (real) descriptions of the benchmarks used. Our results also show that the robustness of LLMs in task description defects depends primarily on the type of defect and the characteristics of the task description, rather than the capacity of the model, with Under-Specification defects being the most severe. We further found that benchmarks with richer contextual grounding, such as LiveCodeBench, exhibit substantially greater resilience, highlighting the importance of structured task descriptions for reliable LLM-based code generation.

\end{abstract}

\keywords{Code generation, LLMs, prompt engineering, Robustness, Benchmarking}

\maketitle

\section{Introduction}
Large language models (LLMs) such as Codex~\cite{chen2021codex}, AlphaCode~\cite{li2022alphacode}, Code Llama~\cite{roziere2023codellama}, and Qwen2.5 Coder~\cite{hui2024qwen25coder}, along with related LLM agents~\cite{openai2025gpt5,anthropic2024claude}, increasingly support developers across a broad spectrum of programming tasks, such as code generation~\cite {generation2026survey}. Previous work demonstrates their practical impact: AI-based coding tools boost developer productivity~\cite{peng2023copilot}.

In parallel, the evaluation methodologies for these systems have evolved substantially. Benchmarks such as HumanEval~\cite{chen2021codex} and MBPP~\cite{mbpp} established the unit-test pass rate as a standard evaluation paradigm. However, subsequent work has revealed important limitations. EvalPlus~\cite{liu2023evalplus} shows that increasing the test coverage significantly reduces the reported performance, highlighting the fragility of previous assessments. More recently, LiveCodeBench~\cite{livecodebench2024} addressed additional concerns by continuously updating problem sets to mitigate data contamination. In addition, it introduces richer problem specifications, incorporating explicit numerical constraints, formal input–output contracts, and illustrative examples alongside natural language descriptions.


Interestingly, the above studies evaluate the generation of code on well-formed and sufficiently described task descriptions. However, in practice, natural language descriptions are inherently imprecise and may be subject to several description issues, similar to specification defects that appear in software specifications \cite{Meyer85,0025377, larbi2025prompts,vogelsang2025requirements}. Therefore, task descriptions can include vague terminology, omit edge-case constraints, or may contain surface-level noise, such as typographical and formatting errors, all of which can drastically affect the correctness of the generated code \cite{larbi2025prompts,SpecFix,zhu2024promptrobust,wu2025humanevalcomm}.  

To address this issue, we developed \textit{SpecValidator}, a simple classifier to automatically detect such task description defects prior to code generation. 
SpecValidator is a lightweight LoRA-fine-tuned predictor trained on a set of artificially injected defects. In particular, we define and apply three defect types — \emph{Lexical Vagueness} (LV), \emph{Under-Specification} (US), and \emph{Syntax and Formatting} (SF) — reflecting the key issues task descriptions may have, based on which we fine-tune our predictor and check its ability to detect them. 

We evaluate 10 LLMs spanning from small (6--7B) to large open-source (15--34B) and state-of-the-art reasoning models across 3 benchmarks and found significant degradation in the models' code correctness. Interestingly, our results show that the structural characteristics of the task descriptions play an important role on the correctness of the generated code. Under-Specification causes the most significant drops, of up to 15.3\%, with no model being robust, while benchmarks with richer contextual grounding, such as LiveCodeBench, are substantially more resilient. Lexical Vagueness produces moderate and benchmark-dependent degradation, whereas surface formatting issues have a negligible effect. 

SpecValidator shows strong defect detection performance with F1 \, = \, 0.804 and MCC \, = \, 0.745, outperforming significantly few-shot GPT-5-mini (F1 \, = \, 0.462 and MCC \, = \, 0.276) and Claude Sonnet~4 (F1 \, = \, 0.557 and MCC \, = \, 0.412). 
Perhaps more importantly, by applying SpecValidator on the original set of task descriptions, we reveal that these surprisingly contain Under-Specification defects (confirmed via thorough manual inspection). This demonstrates the ability of SpecValidator to detect real specification defects. Thus, in a sense we show that SpecValidator generalizes beyond the artificial defects we used for the fine-tuning, with a precision of 72\%, correctly identifying 72\% of the real unseen descriptions as defective.

The main contributions of this paper are:

\begin{itemize}   
   \item A large-scale analysis of 10 models and 3  benchmarks, showing that defects' impact is shaped both by defect type and task description structure. 
  \item \textit{SpecValidator} A lightweight defect detector, a Qwen2.5-Coder-1.5B LoRA fine-tune model, achieving F1\,=\,0.804 and MCC\,=\,0.745, outperforming both GPT-5 mini and Claude Sonnet~4. \textit{SpecValidator} is capable of detecting real description defects in the benchmarks we use.   
  \item A publicly available dataset of 6{,}573 defective task description instances, supporting replication and future research on task description and benchmark design.
\end{itemize}

\section{Related Work}
\subsection{LLM-based code generation and evaluation}

\textit{The HumanEval}~\cite{humaneval}, \textit{MBPP}~\cite{mbpp} and \textit{EvalPlus}~\cite{liu2023evalplus} benchmarks evaluate functional correctness using test pass rates (e.g. \textit{pass@k}). 
\textit{LiveCodeBench}~\cite{livecodebench2024} mitigates benchmark contamination by continuously introducing new competitive-programming problems, while \textit{BigCodeBench}~\cite{zhuo2025bigcodebench}, and Siddiq et al. \cite{siddiq2024fault} extend evaluation to more complex, library-intensive tasks that require diverse API usage and more realistic programming workflows.

Open-weight models include \textit{Code Llama}~\cite{roziere2023codellama}, \textit{StarCoder2}~\cite{lozhkov2024starcoder2}, \textit{DeepSeek-Coder}~\cite{guo2024deepseek}, and \textit{Qwen2.5-Coder}~\cite{hui2024qwen25coder}, while proprietary systems such as GPT-5~\cite{openai2025gpt5} and Claude Code~\cite{anthropic2024claude} represent the state of the art in commercial deployments. \textit{Despite the growing diversity of benchmarks and models, evaluation practices remain largely focused on correctness under well-specified task descriptions, leaving model behavior under realistic description defects unexplored.}

\subsection{Prompt engineering for code tasks}

Li et al.~\cite{li2025scot} proposed Structured Chain-of-Thought (SCoT) prompting, which decomposes generation through program-structure-aware intermediate steps \cite{wei2022chain}, yielding consistent improvements over standard prompting on HumanEval and MBPP. Test-driven feedback loops~\cite{TiCoder} ask users to refine the task description using feedback from tests generated from the original description, using a multi-turn LLM dialog. Self-Debugging~\cite{chen2024selfdebugging} teaches models to self-identify and correct errors in their own outputs via execution-guided feedback. These studies focus on the generation of correct code from well-specified, sufficiently detailed descriptions and excluding potential defects.

\subsection{Detecting description defects}
 
ClarifyGPT~\cite{mu2024clarifygpt} and Larbi et al.~\cite{larbi2025prompts}, and Wu et al. \cite{wu2025humanevalcomm} use LLMs to detect defective task descriptions. These approaches have been evaluated on relatively simple cases, such as HumanEval and MBPP, and on a limited number of defects and types. The baselines used in this work represent the performance of these approaches when using today's LLMs. Jia et al.~\cite{SpecFix} and Tian et al. \cite{tian2025mufix} train models to automatically repair contradictory descriptions. This approach is somewhat orthogonal to our detection one, as it aims only at repairing issues. 
Nevertheless, Jia et al. are using contrastive specification inference, which may not perform well with the under-specified defects we study.

\subsection{LLM robustness on prompt perturbations}

\textit{ReCode}~\cite{recode} introduced over 30 semantic-preserving task description transformations (e.g., variable renaming, dead-code insertion, and docstring reformatting), revealing a substantial but uneven performance degradation of the correctness of the generated code. Mastropaolo et al.~\cite{mastropaolo2023robustness}  showed that semantically equivalent paraphrases of Javadoc can alter the GitHub Copilot output in nearly half of the cases, affecting correctness 28\% of the cases. 

Subsequent work broadened the range of possible perturbations by showing that single token perturbations lead to semantically different code versions \cite{HamidiKP25}. \textit{NLPerturbator}~\cite{NLPerturbator} proposed 18 categories of natural-language variations from real developer data, while Rabbi et al.~\cite{rabbi2025multilanguage} extended the analysis to multiple programming languages. Shirafuji et al. \cite{shirafuji2023exploring} demonstrated that code LLMs are highly sensitive to superficial modifications of problem descriptions. Sclar et al.~\cite{formatspread} highlighted the sensitivity of LLMs to prompt formatting, showing that minor changes such as whitespace or casing can impact code correctness, and PromptRobust\cite{zhu2024promptrobust} showed contemporary LLMs suffer up to 39\% average performance drops from word-level perturbations alone \cite{zhu2024promptrobust,xia2024evoeval}.

The closest work to ours is that of Larbi et al.~\cite{larbi2025prompts}, which found that code correctness is strongly affected by ambiguous, incomplete, and contradictory task descriptions. However, that work did not propose ways to detect such defects.   

Overall, our work goes a step ahead by making a controlled analysis of three defect categories (lexical vagueness, under-specification, and syntax/formatting errors) across 10 models, it develops a lightweight LoRA-based defect classifier evaluated on both artificial (mutants) and real defects. 

\begin{figure*}
    \centering
    \includegraphics[width=0.7\linewidth]{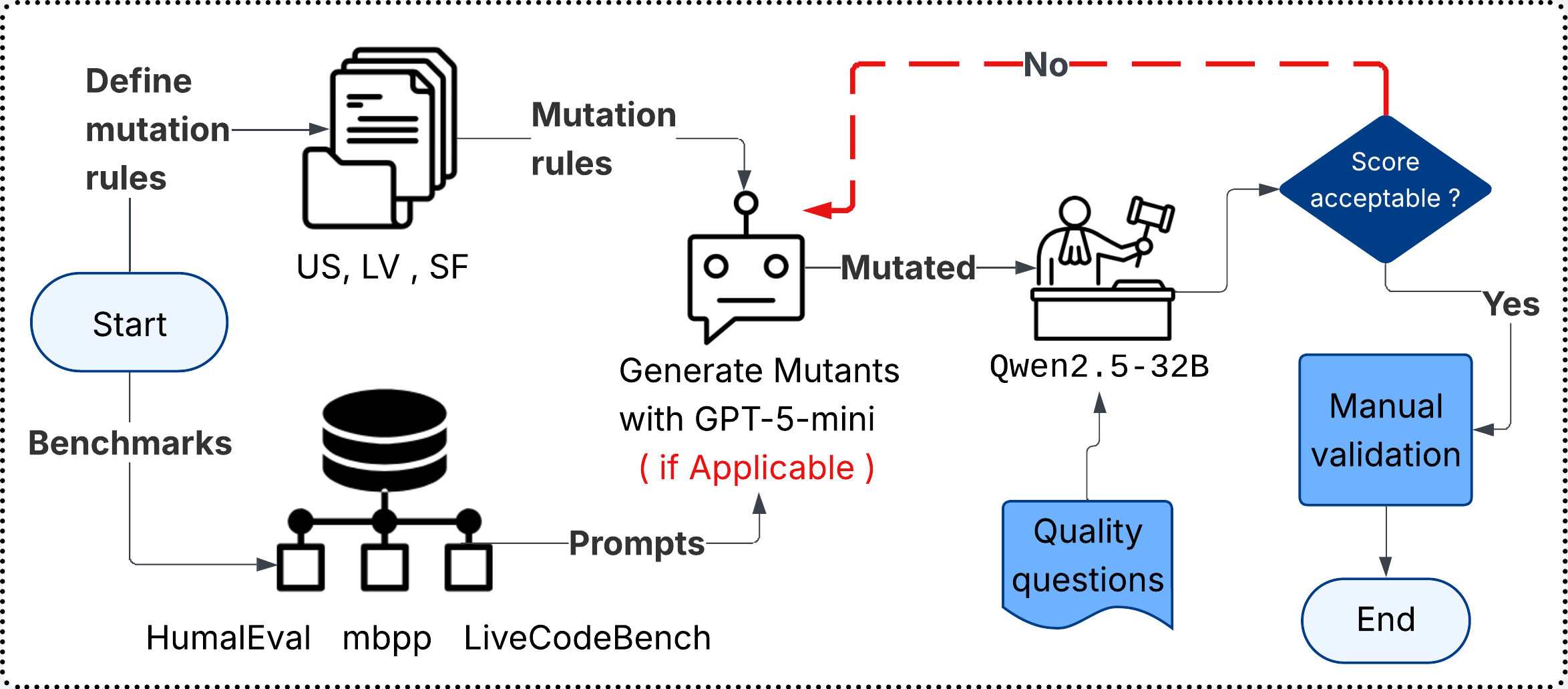}
    \vspace{-0.5em}
    \caption{Dataset construction. Defects are generated with GPT‑5‑mini with defect quality being scored by a Qwen‑32B evaluator. Defective descriptions are then manually verified by 3 human reviewers.}
    \vspace{-0.5em}
    \label{fig:data_generation}
\end{figure*}

\section{Dataset Construction}


 Figure \ref{fig:data_generation} shows the process of dataset construction. Starting from three well-known code-generation benchmarks, we apply three complementary defect-injection strategies to produce our benchmark. Data quality is assessed at two levels: automated evaluation by an independent LLM judge across all 6,573 mutated instances, and manual inspection of a stratified random sample for ground-truth calibration.

\subsection{Benchmarks}

We select three code-generation benchmarks that cover a range of task types, from structured function completion to natural-language descriptions to full programming problems. This variety helps us evaluate model performance across different levels of difficulty, description styles, and reasoning requirements. Table~\ref{tab:benchmarks} records the key characteristics of the benchmarks we use. \textsc{HumanEval}~\cite{humaneval} uses structured docstrings with typed function signatures; \textsc{MBPP}~\cite{mbpp} provides short, informal descriptions in natural-language with minimal scaffolding; and \textsc{LiveCodeBench}~\cite{livecodebench2024} provides competitive-programming problem statements with explicit numerical bounds and formal input/output contracts. All benchmarks are for Python.

\begin{table*}[t]
\centering
\caption{Overview of the three benchmarks used in our experiments   }
\vspace{-0.5em}
\label{tab:benchmarks}
\begin{tabular}{lccccc}
\toprule
\textbf{Benchmark} & \textbf{Size} & \textbf{Task style} & \textbf{Complexity} & \textbf{Language} & \textbf{Evaluation} \\
\midrule
HumanEval
  & 164
  & Function completion (docstring)
  & Easy--Medium
  & Python
  & Unit tests \\
\addlinespace
MBPP
  & 974
  & NL description of short programs
  & Easy
  & Python
  & Unit tests \\
\addlinespace
LiveCodeBench
  & 1{,}055
  & Competitive programming\newline(full problem statement)
  & Easy--Hard
  & Python
  & stdin/stdout \\
\bottomrule
\end{tabular}
\vspace{-0.5em}
\end{table*}
 
\subsection{Generating defective task descriptions}

We construct defective task descriptions by systematically mutating original descriptions. Table \ref {tab:mutation_examples} shows examples of our mutations. We generate three defects per task description (one per defect type). In cases where the original descriptions are too short or have no constraint to delete, no defective variant is produced. Mutations are applied only to task descriptions: reference solutions, test cases, and evaluation harnesses are not modified, so any observed change (e.g. in Pass@1 metric) is solely attributable to the altered description. 

We generate defective descriptions with \texttt{gpt-5-mini} (via the OpenAI Batch). We prompt the model to mutate original descriptions using a predefined set of transformation rules. To ensure that the generated descriptions comply with our requirements, we use an LLM-as-a-judge to assign compliance scores. In case the scores are not above 85\% for all defect types and benchmarks, we manually inspect the low-compliance defective descriptions and refine our prompt to gpt-5-mini until we reach at least this 85\% compliance threshold. All our generation prompts are available in the replication package.\footnote{\url{https://github.com/Amal-AK/detecting_prompt_defects}}. 

\subsubsection{Defect Types}

 Previous work has identified token perturbations and lexical ambiguity as a source of performance variance in code generation models \cite{recode,NLPerturbator,larbi2025prompts}. Missing or under-specified requirements can also cause models to generate code that is syntactically correct but semantically incorrect~\cite{mu2024clarifygpt,SpecFix,TiCoder,HamidiKP25}. Surface-level formatting noise, while seemingly superficial, has been shown to induce performance swings in open-source LLMs~\cite{formatspread,recode}. Together, these three defect classes (lexical vagueness, under-specification, and syntactic/formatting corruption) represent orthogonal failure modes that cover the main ways a natural-language specification defect can appear in the user prompts. Table~\ref{tab:mutation_examples} provides representative examples of each defect type applied to task descriptions from all three benchmarks.

\subsubsection{Mutation Operators}
\paragraph{Lexical Vagueness (LV)}
LV mutations replace precise, task-relevant vocabulary with semantically broader or less informative alternatives, reducing the specificity of the task without removing any explicit constraint. This mirrors the paraphrasing and synonym-substitution 
perturbations studied by Wang et al.~\cite{recode} and Chen et al.~\cite{NLPerturbator}. Concretely, an LV mutation consists of one to five of the following transformations:

\begin{itemize}
  \item Replace domain-specific terms with broader synonyms 
        (e.g., \textit{sort} $\rightarrow$ \textit{arrange}).
  \item Weaken action verbs to less precise ones 
        (e.g., \textit{insert} $\rightarrow$ \textit{put}).
  \item Rename function and parameter identifiers with less informative synonyms 
        (e.g., \texttt{delimiter} $\rightarrow$ \texttt{filler}).
  \item Generalize or remove type annotations.
  \item Make multiple lexical changes, compounding the ambiguity.
\end{itemize}

\paragraph{Under-Specification (US)}
US mutations remove one explicit constraint from the task description, producing a description that remains grammatically well-formed but is semantically under-specified. This defect type is motivated by empirical findings that incomplete requirements cause most of the intent mismatch in LLM-based code generation~\cite{larbi2025prompts}.  To ensure under-specified defects, we restrict deletion to the following elements of the description:

\begin{itemize}
  \item Numeric bounds or threshold values.
  \item Ordering or sorting rules.
  \item Input preconditions or domain restrictions.
  \item Output edge-case behavior.
  \item Formula or behavioral pinning constraints.
  \item Error or exception handling requirements.
  \item Output type or format constraints.
\end{itemize}

\paragraph{Syntax and Formatting (SF)}
SF mutations introduce surface-level noise into the task description without altering its semantic content, targeting the susceptibility of LLMs to low-level textual  corruption~\cite{formatspread}. Sclar et al.~\cite{formatspread} demonstrated that 
format variations alone can have a significant impact, motivating this defect type. Each SF mutation applies one of the following corruption types:

\begin{itemize}
  \item Token-level corruption (character transposition, typos).
  \item Delimiter, bracket, or colon corruption.
  \item Indentation or whitespace layout corruption.
  \item Example-block formatting corruption.
\end{itemize}


\subsection{Defect validation}

We performed an automated assessment of the generated defects using an independent LLM judge followed by manual analysis of a stratified random sample.


\textbf{Judge model:} 
We select \textit{Qwen2.5-Coder-32B-Instruct}~\cite{hui2024qwen25codertechnicalreport} 
as the judge model. Three properties motivate this choice: (i)~it is a strong open-source, code-specialized model that is categorically different from the generation model (\texttt{gpt-5-mini}), reducing the risk of generator-judge collusion common in LLM-as-a-judge setups~\cite{llmjudge}; (ii)~it is deployed locally on our own infrastructure, ensuring full reproducibility without dependence on external APIs; and (iii)~its instruction following capability is sufficient for the structured binary-judgment protocol described below. The model is run at float16 precision with greedy decoding, and its output is constrained to a single JSON object (\texttt{\{"score": 0\}} or \texttt{\{"score": 1\}}).

\textbf{Defect evaluation criteria:} 
All defective descriptions are evaluated on the basis of the following criteria: 

\begin{itemize}
   \item LV compliance: \textit{Did the mutation replace specific terms, variable names, or descriptions with vaguer counterparts without changing the overall task?}
  
   \item SF compliance: \textit{Does the defective description contain only typographical or formatting errors compared to the original?}

   \item US compliance: \textit{Is exactly one requirement, condition, or detail missing in the defective description compared to the original one?}

   \item Naturalness: \textit{Does the defective description read like a natural description a real user could write?}

\end{itemize}

\textbf{Defect evaluation results:}
Table~\ref{tab:judge_quality} reports compliance and naturalness scores for each benchmark–defect pair. The LV and SF mutations achieve near-perfect compliance (0.98–1.00) and high naturalness (0.96–1.00) in all data. US shows lower compliance (0.85–0.89), mainly due to structural constraints in shorter MBPP prompts and stricter specification structures in LiveCodeBench. Naturalness remains high for LV and SF, but drops for US on LiveCodeBench (0.61), where removing constraints can produce unusually terse problem statements.

\begin{table}[t]
\centering
\vspace{-0.1em}
\caption{Evaluating mutations by judge (Qwen2.5-Coder-32B-Instruct). All values are proportions 
(0--1).}
\vspace{-0.6em}
\label{tab:judge_quality}
\resizebox{\columnwidth}{!}{%
\begin{tabular}{llrcc}
\toprule
\textbf{Dataset} & \textbf{Mutation} & \textbf{$N$} 
  & \textbf{Compliance} & \textbf{Naturalness} \\
\midrule
\multirow{3}{*}{HumanEval}
  & LV & 163  & 0.99 & 0.99 \\
  & SF & 164  & 1.00 & 0.99 \\
  & US & 160  & 0.89 & 0.81 \\
\midrule
\multirow{3}{*}{MBPP}
  & LV & 974  & 1.00 & 0.99 \\
  & SF & 974  & 0.98 & 1.00 \\
  & US & 974  & 0.88 & 0.82 \\
\midrule
\multirow{3}{*}{LiveCodeBench}
  & LV & 1054 & 1.00 & 0.96 \\
  & SF & 1055 & 1.00 & 0.99 \\
  & US & 1055 & 0.85 & 0.61 \\
\bottomrule
\end{tabular}%
}
\vspace{-0.7em}
\end{table}

\textbf{Manual Validation:} To assess the reliability of the automated judge, we manually annotated a stratified random sample of 100 mutants of all types (LV, US, SF) and benchmarks (HumanEval, MBPP, LiveCodeBench), sampling roughly 11 samples per benchmark–mutation pair. Three researchers independently evaluated each instance using the same compliance and naturalness criteria as the LLM judge, with disagreements resolved by majority vote. Human annotators rated 86\% of instances as natural and 97\% as compliant, closely matching the judge’s assessments.

\begin{table*}[ht]
\centering
\small
\caption{Representative examples of the three mutation strategies.
\textbf{US}: removes a constraint; \textbf{LV}: paraphrases vocabulary
and renames identifiers; \textbf{SF}: introduces typographical and
formatting noise.}
\vspace{-0.3em}
\label{tab:mutation_examples}
\renewcommand{\arraystretch}{1.2}
\setlength{\tabcolsep}{4pt}
\begin{tabular}{p{1.8cm} p{\dimexpr\linewidth-1.8cm-2\tabcolsep\relax}}
\toprule
\textbf{Mutation} & \textbf{Prompt} \\
\midrule
\multicolumn{2}{l}{\textit{HumanEval / 5}} \\[2pt]
Original  & \texttt{def intersperse(numbers: List[int], delimeter: int) -> List[int]:}
            \textit{``Insert a number `delimeter' \underline{between every two consecutive elements} of `numbers'.''}
            \texttt{>>> intersperse([1,2,3], 4)} $\Rightarrow$ \texttt{[1,4,2,4,3]} \\[3pt]
US        & \texttt{def intersperse(numbers: List[int], delimeter: int) -> List[int]:}
            \textit{``Insert a number `delimeter' \underline{into} the input list `numbers'.''}
            \texttt{>>> intersperse([1,2,3], 4)} $\Rightarrow$ \texttt{[1,4,2,4,3]} \\[3pt]
LV        & \texttt{def \underline{place\_between}(\underline{items}, \underline{filler}):}
            \textit{``Put a value `filler' between \underline{neighboring entries} of the \underline{sequence}.''}
            \texttt{>>> place\_between([1,2,3], 4)} $\Rightarrow$ \texttt{[1,4,2,4,3]} \\[3pt]
SF        & \texttt{def intersperse(numbers: List[int\underline{)}, \underline{delimter}: int) -> List[int\underline{)}}
            \texttt{\underline{'''} Insert a number `\underline{delimter}' between every two consecutive elements \ldots} \\
\midrule
\multicolumn{2}{l}{\textit{MBPP / 12}} \\[2pt]
Original  & \textit{``Write a function to sort a given matrix in \underline{ascending order} according to the sum of its rows.''} \\[3pt]
US        & \textit{``Write a function to sort a given matrix according to the sum of its rows.''} \\[3pt]
LV        & \textit{``Write a function to \underline{arrange} a given matrix in ascending order according to the \underline{total} of its rows.''} \\[3pt]
SF        & \textit{``Write a function to sort a given matrix in \underline{ascdcending} order according to the sum of its rows.''} \\
\midrule
\multicolumn{2}{l}{\textit{LiveCodeBench / 2811}} \\[2pt]
Original  & \textit{``An array of \underline{distinct positive} integers is called a k-avoiding array if no pair sums to $k$. Return the minimum possible sum of length $n$.''}
            \texttt{n=5,k=4}$\Rightarrow$\texttt{18} \; \texttt{n=2,k=6}$\Rightarrow$\texttt{3} \\[3pt]
US        & \textit{``An array of integers is called a k-avoiding array if no pair sums to $k$. Return the minimum possible sum of length $n$.''}
            \texttt{n=5,k=4}$\Rightarrow$\texttt{18} \; \texttt{n=2,k=6}$\Rightarrow$\texttt{3} \\[3pt]
LV        & \textit{``A \underline{list} of \underline{different positive numbers} where no two \underline{add up to} $k$. Return the \underline{smallest total} of length $n$.''}
            \texttt{n=5,k=4}$\Rightarrow$\texttt{18} \; \texttt{n=2,k=6}$\Rightarrow$\texttt{3} \\[3pt]
SF        & \texttt{\underline{INPUT;\{}You are given two \underline{integres} $n$ and $k$ \ldots \underline{**INVALID****OCUMENT**\}**}} \\
\bottomrule
\end{tabular}
\end{table*}
\section{Study Design}

\subsection{Research questions}

Developers typically write imperfect descriptions: they may use imprecise vocabulary, omit edge cases, or write noisy statements. The robustness of code LLMs to such issues remains underexplored and there is also no systematic infrastructure to detect or mitigate these defects. Our work explores the following questions:

\textbf{RQ1 (Impact of defects): How lexical vagueness, under-specification, and syntactic noise affect generated code correctness across benchmarks of varying complexity?}

To answer this question, we evaluate 10 code-generation LLMs covering 3 model sizes: 3 small models (6–7B), 5 medium-to-large models (15–34B), and 2 reasoning models (GPT-5-mini and Claude Sonnet 4). We conduct experiments on 3 widely used benchmarks: HumanEval, MBPP, and LiveCodeBench. For each benchmark, we introduce 3 types of defect (US, LV, SF). We measure the performance of each model in generating correct code using Pass@1 metric. We examine to what extent robustness varies with model scale, benchmark complexity, and defect type.

\textbf{RQ2 (Defect detection): Can a small fine-tuned model detect description defects as well as large model baselines?}   

We develop SpecValidator, an approach based on small, locally deployable classifier to detect dejects in code task descriptions. We aim to assess whether SpecValidator can match the performance of frontier models on four-class defect detection (clean, LV, US, SF). For the small model, we fine-tune Qwen2.5-Coder-1.5B under three regimes of increasing parameter budget: linear probing, LoRA, and full fine-tuning. We compare it with GPT-5-mini and Claude Sonnet 4 queried with zero shot and few shots.

\textbf{RQ3 (Defect detection on original prompts): Does SpecValidator detect (unseen) real-world defects?}
We want to assess whether SpecValidator can generalize to unseen defects in real-world specifications, beyond the synthetic defects we previously generated. To this end, we apply SpecValidator on the original descriptions of our benchmarks. and examine whether flagged instances reflect genuine defects in the ground truth descriptions.

Once a defect is detected, its type is reported to the user in order to understand the issue with the description and update it accordingly.


\subsection{Defect detection and SpecValidator}

We classify task descriptions in one of four categories: \textsc{Clean}, \textsc{LV}, \textsc{US}, or \textsc{SF}. We evaluate two families of approaches.

\textbf{Proprietary LLM systems.} We query GPT-5-mini and Claude Sonnet 4 with a structured prompt that presents the code generation task description and instructs the model to identify the defect type, if any. In addition to the zero-shot setting, we also evaluate a few-shot variant \cite{brown2020language} where the model is provided with four labeled examples prior to classification. These serve as strong baselines that do not require training data and leverage the broad knowledge of frontier models.

\begin{tcolorbox}[
  colback=gray!5,
  colframe=gray!40,
  boxrule=0.4pt,
  fontupper=\scriptsize,
  left=6pt, right=6pt, top=4pt, bottom=4pt,
  title=The prompt used for Zero-Shot classification. (4 Examples are added in case of few-shot) 
]
{\small\ttfamily
You are a code-benchmark quality auditor.
Classify the following coding prompt into exactly one of:\\[4pt]
\hspace*{1em}\textbf{LV}\hspace{1em} -- The prompt uses vague or imprecise wording.\\
\hspace*{1em}\textbf{SF}\hspace{1em} -- The prompt contains syntax or formatting errors.\\
\hspace*{1em}\textbf{US}\hspace{1em} -- The prompt is missing a constraint or condition.\\
\hspace*{1em}\textbf{CLEAN} -- The prompt is complete and well-formed.
}
\end{tcolorbox}

\textbf{SpecValidator (fine-tuned classifier)} We fine-tune Qwen2.5-Coder-1.5B \cite{hui2024qwen25coder} as a sequence classifier on a labeled dataset constructed from our mutation approach. Each example consists of a task description (original or mutated) paired with its defect label. We train in three parameter budgets to assess the cost–accuracy trade-off: (i)~\textit{linear probing}, where only the classification head is trained and the backbone is frozen; (ii)~\textit{LoRA adaptation}, where low-rank updates are applied to attention projection layers as described in the following implementation section; and (iii)~\textit{full fine-tuning}, where all parameters are updated.

The training set is composed of equal numbers of original and mutated description (8,766 samples in total). The dataset is split 80/10/10 into train, validation, and test sets, stratified by label. The finetuning is performed twice with two different seeds.

\subsection{Selected LLMs}

To study the impact of defects, we choose a diverse set of code generation models to ensure that our findings are not artifacts of a single architecture or training regime. Models are grouped into the following three categories/families:

\begin{itemize}   
   \item  \textbf{Small open-source models} ($\sim$7B parameters): Qwen2.5-Coder-7B~\cite{hui2024qwen25coder}, DeepSeek-Coder-6.7B~\cite{guo2024deepseek}, and CodeLlama-7B~\cite{roziere2023codellama}. These models are representative of deployable, resource-constrained settings and are widely used as baselines by the literature.

   \item  \textbf{Large open-source models} (15B--34B parameters): Qwen2.5-Coder-32B~\cite{hui2024qwen25coder}, DeepSeek-Coder-33B~\cite{guo2024deepseek}, CodeLlama-34B~\cite{roziere2023codellama}, Codestral-22B~\cite{mistral2024codestral}, and StarCoder2-15B~\cite{lozhkov2024starcoder2}. This tier allows us to examine whether larger capacity yields greater robustness to prompt mutations, independent of proprietary training data or alignment procedures.

   \item  \textbf{Closed-source LLM-based systems}:  GPT-5 mini~\cite{openai2025gpt5}, and Claude Sonnet 4~\cite{anthropic2024claude}. These represent the current state-of-the-art in code generation and serve as upper-bound references. Their inclusion enables a direct comparison between frontier models and open-source alternatives under identical evaluation conditions.
\end{itemize} 

Collectively, this selection spans five model families, two provider categories (open-source and proprietary), and two scale regimes, enabling controlled cross-family and cross-scale comparisons. 

\subsection{Implementation }
\label{{impl}}
\subsubsection{\textbf{Metrics}}
Pass@1~\cite{chen2021codex} is used as the primary correctness metric, as it reflects the real-world scenario where a developer accepts the model's first suggestion. This metric is the de facto standard in code generation evaluation~\cite{chen2021codex, mbpp} and enables a direct comparison with previous work.

For the four-class mutation classification task (LV, SF, US, CLEAN), we report the Matthews Correlation Coefficient (MCC)~\cite{matthews1975mcc},  macro-F1, precision, recall, and accuracy. MCC is preferred over accuracy when there is class imbalance as it accounts for all entries in the confusion matrix.

\subsubsection{\textbf{Technical Details}}
\textbf{Inference.} All models are run with greedy decoding and temperature $= 0$), ensuring that the outputs are theoretically deterministic given a fixed prompt. Under this setting, each model produces exactly one outcome per problem, and re-running the same experiment yields identical results. \\
The reasoning models (GPT and Claude) are queried via the OpenAI and Anthropic batch APIs, while open-source models are served locally on NVIDIA A100 GPUs in float16 precision. Generated solutions are extracted via markdown fenced code block parsing, with fallback to model-specific delimiters (e.g., CodeLlama's \texttt{[PYTHON]} tags). Each solution is executed in an isolated subprocess with a 20-second timeout. The code and data are available. \footnote{Replication package: 
\url{https://github.com/Amal-AK/detecting_prompt_defects}}

\textbf{LoRA fine-tuning.} We apply LoRA~\cite{hu2022lora} to the attention projection layers with rank $r{=}16$ and $\alpha{=}32$, which prior work has identified as an effective configuration for code model adaptation~\cite{hu2022lora,dettmers2023qlora,houlsby2019parameter}. Training uses AdamW~\cite{loshchilov2019adamw} with learning rate $2\times10^{-4}$, weight decay $10^{-4}$, cosine decay scheduling, and 100 warmup steps. An effective batch size of 16 is achieved via gradient accumulation. Early stopping (patience = 4) is applied based on the validation performance to prevent overfitting. All finetuning experiments use both seeds 42 and 123456; the results are averaged.


\section{Experimental Results}

\subsection{Impact of defects on code generation}

Table~\ref{tab:all_results} summarizes the evaluation of ten code LLMs on three benchmarks (HumanEval, MBPP, and LiveCodeBench) under three types of mutation: Under-Specification (US), Lexical Vagueness (LV), and Syntactic/Formatting errors (SF). 
The results show that these mutation types induce substantially different levels of performance degradation in models and benchmarks, suggesting that description defects vary in their impact and should not be treated uniformly.

\begin{table*}[t]
\centering
\vspace{0.2em}
\caption{
  Pass@1 results on all benchmarks and mutation types.
  For each mutation (US, LV, SF), Pass@1 deltas wrt to the original
  are shown in {\color{red}red} ($\downarrow$\,drop) and
  {\color{green!60!black}green} ($\uparrow$\,gain).
  \textbf{US} = Under-Specification;
  \textbf{LV} = Lexical Vagueness;
  \textbf{SF} = Syntax and Formatting.
}
\label{tab:all_results}
\resizebox{\textwidth}{!}{%
\begin{tabular}{l|rrrr|rrrr|rrrr}
\toprule
& \multicolumn{4}{c|}{\textbf{HumanEval}}
& \multicolumn{4}{c|}{\textbf{MBPP}}
& \multicolumn{4}{c}{\textbf{LiveCodeBench}} \\
\cmidrule(lr){2-5}\cmidrule(lr){6-9}\cmidrule(lr){10-13}
\textbf{Model}
  & \textbf{Orig} & \textbf{US} & \textbf{LV} & \textbf{SF}
  & \textbf{Orig} & \textbf{US} & \textbf{LV} & \textbf{SF}
  & \textbf{Orig} & \textbf{US} & \textbf{LV} & \textbf{SF} \\
\midrule
\multicolumn{13}{l}{\textit{Small models ($\leq$7B)}} \\[2pt]
CodeLlama-7B
  & 37.2 & 29.3\,{\color{red}$\downarrow$7.9}  & 29.3\,{\color{red}$\downarrow$7.9}  & 37.2\,{\color{gray}=}
  & 33.3 & 26.8\,{\color{red}$\downarrow$6.5}  & 31.2\,{\color{red}$\downarrow$2.1}  & 33.7\,{\color{green!60!black}$\uparrow$0.4}
  &  8.1 &  8.4\,{\color{green!60!black}$\uparrow$0.3}  &  8.2\,{\color{green!60!black}$\uparrow$0.1}  &  7.8\,{\color{red}$\downarrow$0.3} \\

DeepSeek-Coder-6.7B
  & 72.6 & 57.3\,{\color{red}$\downarrow$15.3} & 60.4\,{\color{red}$\downarrow$12.2} & 68.9\,{\color{red}$\downarrow$3.7}
  & 44.1 & 36.2\,{\color{red}$\downarrow$7.9}  & 41.6\,{\color{red}$\downarrow$2.5}  & 43.5\,{\color{red}$\downarrow$0.6}
  & 16.3 & 16.0\,{\color{red}$\downarrow$0.3}  & 14.8\,{\color{red}$\downarrow$1.5}  & 15.2\,{\color{red}$\downarrow$1.1} \\

Qwen2.5-Coder-7B
  & 82.3 & 67.7\,{\color{red}$\downarrow$14.6} & 75.6\,{\color{red}$\downarrow$6.7}  & 81.7\,{\color{red}$\downarrow$0.6}
  & 50.1 & 41.2\,{\color{red}$\downarrow$8.9}  & 48.6\,{\color{red}$\downarrow$1.5}  & 49.0\,{\color{red}$\downarrow$1.1}
  & 20.5 & 20.3\,{\color{red}$\downarrow$0.2}  & 22.2\,{\color{green!60!black}$\uparrow$1.7}  & 20.5\,{\color{gray}=} \\
\midrule
\multicolumn{13}{l}{\textit{Large models (15B--34B)}} \\[2pt]
StarCoder2-15B
  & 65.9 & 51.2\,{\color{red}$\downarrow$14.7} & 54.3\,{\color{red}$\downarrow$11.6} & 62.8\,{\color{red}$\downarrow$3.1}
  & 42.1 & 35.3\,{\color{red}$\downarrow$6.8}  & 39.7\,{\color{red}$\downarrow$2.4}  & 42.1\,{\color{gray}=}
  &  6.8 &  8.0\,{\color{green!60!black}$\uparrow$1.2}  &  7.3\,{\color{green!60!black}$\uparrow$0.5}  &  7.6\,{\color{green!60!black}$\uparrow$0.8} \\

Codestral-22B
  & 72.6 & 58.5\,{\color{red}$\downarrow$14.1} & 65.2\,{\color{red}$\downarrow$7.4}  & 70.7\,{\color{red}$\downarrow$1.9}
  & 47.6 & 38.9\,{\color{red}$\downarrow$8.7}  & 45.0\,{\color{red}$\downarrow$2.6}  & 47.7\,{\color{green!60!black}$\uparrow$0.1}
  & 23.0 & 22.5\,{\color{red}$\downarrow$0.5}  & 22.7\,{\color{red}$\downarrow$0.3}  & 22.6\,{\color{red}$\downarrow$0.4} \\

CodeLlama-34B
  & 51.2 & 42.1\,{\color{red}$\downarrow$9.1}  & 45.1\,{\color{red}$\downarrow$6.1}  & 50.0\,{\color{red}$\downarrow$1.2}
  & 37.9 & 29.7\,{\color{red}$\downarrow$8.2}  & 34.7\,{\color{red}$\downarrow$3.2}  & 36.3\,{\color{red}$\downarrow$1.6}
  & 13.6 & 11.6\,{\color{red}$\downarrow$2.0}  & 13.1\,{\color{red}$\downarrow$0.5}  & 12.4\,{\color{red}$\downarrow$1.2} \\

DeepSeek-Coder-33B
  & 72.0 & 61.6\,{\color{red}$\downarrow$10.4} & 68.3\,{\color{red}$\downarrow$3.7}  & 73.8\,{\color{green!60!black}$\uparrow$1.8}
  & 47.0 & 39.8\,{\color{red}$\downarrow$7.2}  & 45.4\,{\color{red}$\downarrow$1.6}  & 46.4\,{\color{red}$\downarrow$0.6}
  & 20.3 & 19.4\,{\color{red}$\downarrow$0.7}  & 19.9\,{\color{red}$\downarrow$0.4}  & 18.8\,{\color{red}$\downarrow$1.5} \\

Qwen2.5-Coder-32B
  & 85.4 & 73.2\,{\color{red}$\downarrow$12.2} & 84.1\,{\color{red}$\downarrow$1.3}  & 86.6\,{\color{green!60!black}$\uparrow$1.2}
  & 51.6 & 40.8\,{\color{red}$\downarrow$10.8} & 48.6\,{\color{red}$\downarrow$3.0}  & 51.4\,{\color{red}$\downarrow$0.2}
  & 32.0 & 30.6\,{\color{red}$\downarrow$1.4}  & 31.7\,{\color{red}$\downarrow$0.3}  & 31.2\,{\color{red}$\downarrow$0.8} \\
\midrule
\multicolumn{13}{l}{\textit{Reasoning models (API)}} \\[2pt]
GPT-5-mini
  & 96.3 & 86.6\,{\color{red}$\downarrow$9.7}  & 89.0\,{\color{red}$\downarrow$7.3}  & 93.3\,{\color{red}$\downarrow$3.0}
  & 49.7 & 44.6\,{\color{red}$\downarrow$5.1}  & 46.9\,{\color{red}$\downarrow$2.8}  & 49.5\,{\color{red}$\downarrow$0.2}
  & 52.5 & 48.5\,{\color{red}$\downarrow$4.0}  & 52.5\,{\color{gray}=}               & 53.2\,{\color{green!60!black}$\uparrow$0.7} \\

Claude Sonnet 4
  & 95.7 & 86.0\,{\color{red}$\downarrow$9.7}  & 89.0\,{\color{red}$\downarrow$6.7}  & 95.7\,{\color{gray}=}
  & 53.0 & 44.1\,{\color{red}$\downarrow$8.9}  & 50.3\,{\color{red}$\downarrow$2.7}  & 53.5\,{\color{green!60!black}$\uparrow$0.5}
  & 51.1 & 49.6\,{\color{red}$\downarrow$1.5}  & 49.9\,{\color{red}$\downarrow$1.2}  & 50.7\,{\color{red}$\downarrow$0.4} \\
\bottomrule
\end{tabular}
}
\vspace{0.2em}
\end{table*}

\begin{figure*}[t]
    \centering
    \vspace{-0.5em}
    \includegraphics[width=0.75\linewidth]{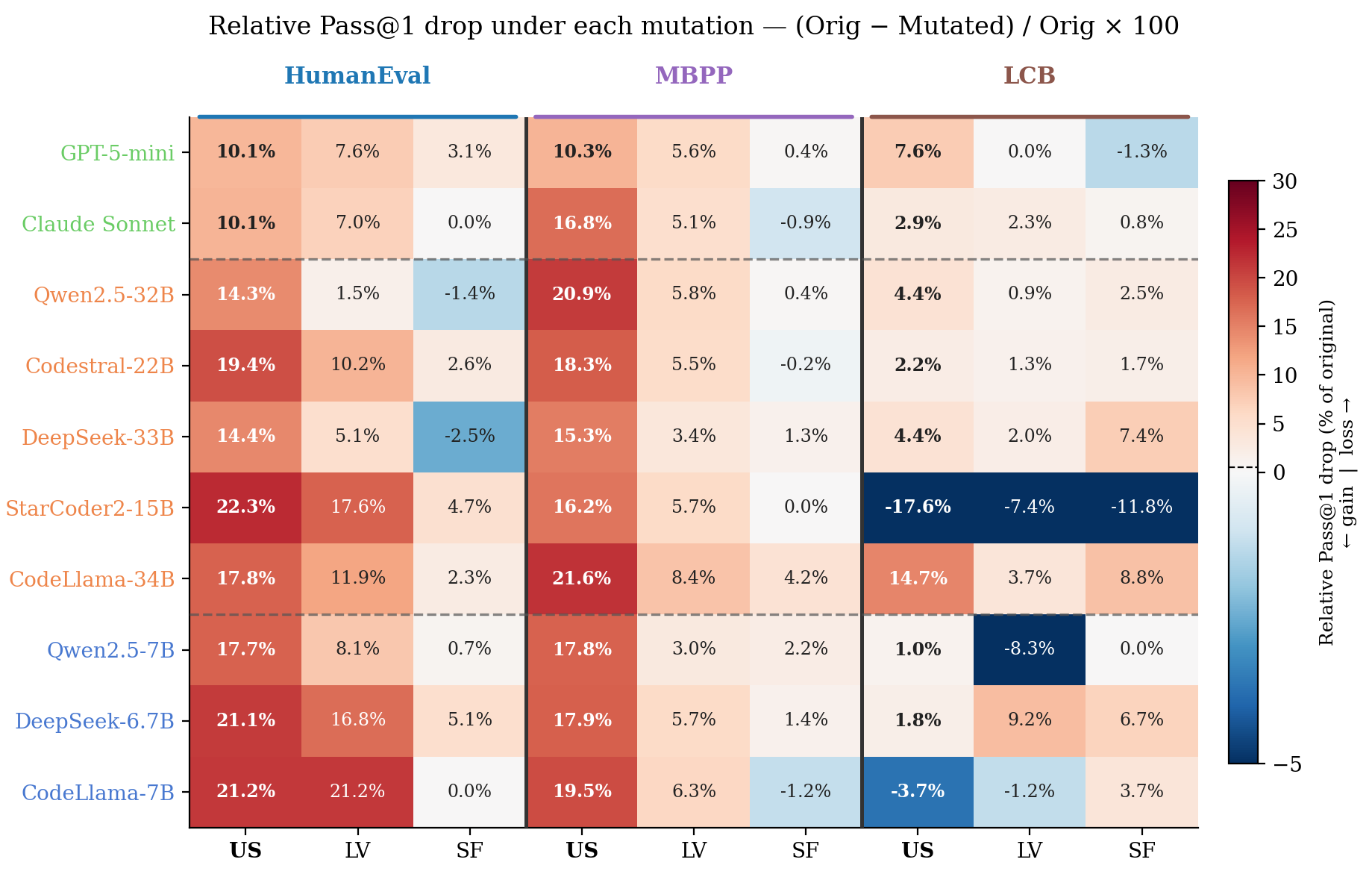}
     \caption{Pass@1 drop (\%) with all defect types in all models and benchmarks, computed as $(\text{Orig} -\text{Mutated}) / \text{Orig} \times 100$. Red cells indicate performance degradation; blue cells indicate a slight gain. Column labels: \textbf{US} = Under-Specification; \textbf{LV} = Lexical Vagueness; \textbf{SF} = Syntax and Formatting.}
    \label{fig:rq1_heatmap}
    \vspace{-0.2em}
\end{figure*}

\textbf{Under-specification is consistently the most damaging mutation.} 
As shown in Table~\ref{tab:all_results}, US mutations cause Pass@1 drops of 7.9--15.3\% on HumanEval across all model scales. Even state-of-the-art reasoning models are substantially affected: GPT-5-mini and Claude~Sonnet~4, despite baselines of 96.3\% and 95.7\% respectively, each decline by 9.7\%. A consistent pattern holds on MBPP, where US induces drops of 5.1--10.8\% across all tiers.

Figure~\ref{fig:rq1_heatmap} confirms that the US column is mostly red in both benchmarks and in all model families, with no model showing immunity. The uniformity of this degradation across architectures and scales indicates that the cause is intrinsic to the mutation itself: removing a constraint leaves the specification compatible with multiple plausible implementations, most of which are incorrect.

\textbf{Lexical vagueness produces benchmark-dependent, moderate drops.} 
On HumanEval, LV mutations produce drops ranging from 1.3\% (Qwen2.5-Coder-32B) to 12.2\% (DeepSeek-Coder-6.7B), with CodeLlama-7B and StarCoder2-15B losing 7.9\% and 11.6\%, respectively (Table~\ref{tab:all_results}).
The nearly negligible drop for Qwen2.5-Coder-32B suggests that larger Qwen-family models are better calibrated to lexical synonymy. On MBPP, however, LV-induced drops shrink markedly across all tiers, ranging from only 1.5\% to 3.2\%, indicating that the shorter and more informal task description style of MBPP limits the impact of vocabulary-level ambiguity. Figure~\ref{fig:rq1_heatmap} makes this benchmark dependence visually apparent: the LV column shows a mixed pattern on HumanEval but near-uniform light shading on MBPP.

\textbf{Surface-level formatting noise has minimal impact.}
SF mutations produce drops below 3.7\% on HumanEval and below 1.6\% on MBPP for virtually all models (Table~\ref{tab:all_results}), with several models showing marginal gains (e.g., DeepSeek-Coder-33B: $+$1.8\% on HumanEval; Qwen2.5-Coder-32B: $+$1.2\%). The SF column in Figure~\ref{fig:rq1_heatmap} is the lightest across all benchmarks, visually confirming that typographical noise is the weakest of the three mutation types. This suggests that code-specialized LLMs are robust to surface-level corruption, extending the observations of Sclar~et~al.~\cite{formatspread}, from general-purpose LLMs to code generation ones.

\textbf{LiveCodeBench is nearly immune to all mutation types. } 
In all models, Pass@1 deltas on LCB remain within $\pm$2.0\% for all defect types (Table~\ref{tab:all_results}), with many models recording marginal gains under US (e.g., StarCoder2-15B: $+$1.2\%; CodeLlama-7B: $+$0.3\%). Figure~\ref{fig:rq1_heatmap} confirms this near-zero sensitivity: the LCB rows are visually the least affected, in sharp contrast to HumanEval and MBPP.
This is explained by the benchmark design: LCB problems include explicit sample input/output examples, so models can reconstruct the correct solution from the examples alone. US mutations remove one textual constraint while the sample I/O, which often encodes the same constraint implicitly, remains intact.
Consequently, we find that descriptions with richer context leads to more robustness in code generation, highlighting an important boundary condition for the generality of our findings.

\textbf{Model size does not relate to robustness; sensitivity is description-dependent.}
 Comparison of small and large models shows that model size does not relate to model robustness. GPT-5-mini and Claude Sonnet 4 lose nearly 10\% on code correctness due to US defects. Sensitivity appears to be description-dependent rather than capacity-dependent: Figure 2 shows different decreases in Pass@1 across all pairs of model-defect types, suggesting that model size alone does not determine a clear trend.

\begin{mdframed}[style=takeawaystyle]
\textbf{RQ1 Takeaway.}
Not all defect types harm code generation equally:
Under-Specification degrades performance by up to 15\% on HumanEval and MBPP, LV causes moderate losses, and SF is negligible. Benchmarks with richer, more realistic descriptions, such as LiveCodeBench, are inherently more resilient to description defects.  Model size and architecture alone do not determine robustness. 
\end{mdframed}

\subsection{Defect detection}

The RQ1 results suggest that the robustness to defects is primarily determined by the characteristics of the input description, rather than the underlying models. Consequently, improving the correctness of code generation requires identifying and addressing defects before code generation.

\begin{figure*}[ht]
  \centering
  \includegraphics[width=\textwidth]{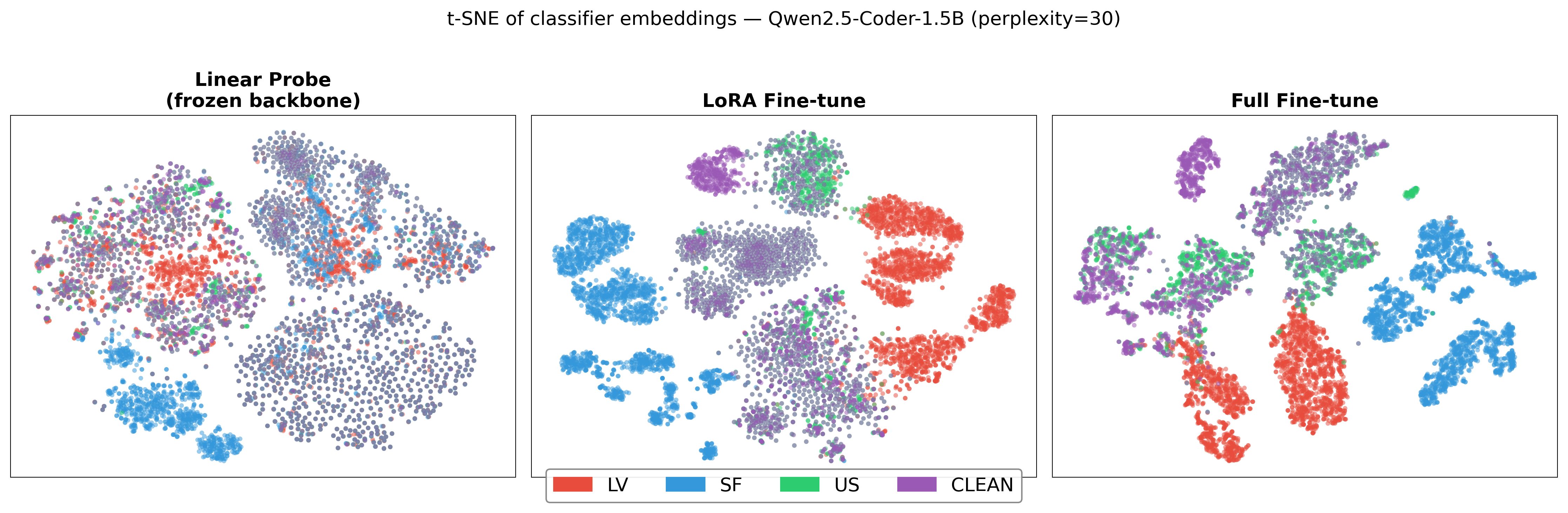}
  \caption{t-SNE projections \cite{vandermaaten2008tsne} of the embedding space of the classifiers (Qwen2.5-Coder-1.5B backbone). From left to right: Linear Probe (frozen backbone), LoRA fine-tune, and Full fine-tune. 
  }
  \label{fig:tsne_classifiers}
\end{figure*}

Table~\ref{tab:classifier_results} reports the classification performance of all five approaches, and Figure~\ref{fig:tsne_classifiers} visualizes the
embedding space learned under each fine-tuning regime.

\textbf{Zero-shot frontier models perform poorly in defect classification}. GPT-5-mini achieves an F1 of 0.469 and an MCC of 0.281, barely above the chance for a four-class problem. Claude Sonnet 4 does slightly better (F1=0.518, MCC=0.359) but still falls well short of practical utility. This result is notable because both models achieve the highest Pass@1 scores in RQ1; strong code generation does not transfer to meta-reasoning about description quality in a zero-shot regime.

\textbf{Few-shot prompting provides only limited gains for frontier models}. Supplying four in-context examples yields marginal improvements for Claude Sonnet 4 (F1=0.557, MCC=0.412), while GPT-5-mini shows no consistent benefit and slightly degrades overall performance (F1 = 0.462, MCC = 0.276). This suggests that, unlike generation tasks, defect classification is not easily improved through in-context learning alone. 

\textbf{Even a frozen linear probe substantially outperforms few-shot frontier models}. Training only a linear classification head on frozen Qwen2.5-Coder-1.5B embeddings yields F1=0.748 and MCC=0.665, a jump of 0.23\% F1 over Claude Sonnet 4 while updating just 0.0004\% of parameters. 

\textbf{SpecValidator with LoRA fine-tuning achieves the best classification performance, outperforming full fine-tuning}. With only 0.14\% of the parameters updated, LoRA achieves F1=0.804 and MCC=0.745, the best results across all classifiers. In contrast, complete fine-tuning (100\% parameters) performs slightly worse (F1=0.788, MCC=0.722). The cause is likely overfitting: the dataset, while containing 8766 instances, is still constrained in size, and full fine-tuning may degrade the backbone's generalization across the input distribution. This result advocates for LoRA as the practical choice: superior accuracy, dramatically lower computations, and local deployability.

\textbf{The ambiguity between clean and under-specified descriptions is the dominant error mode}. 
Unlike LV and SF mutations, which introduce detectable surface signals, US mutations remove a constraint without leaving any positive textual trace: the resulting defective description is grammatically well-formed and stylistically indistinguishable from a clean one. This makes the clean/US boundary structurally harder to learn, as the
classifier must detect the \emph{absence} of information rather than the presence of an anomaly.
Figure~\ref{fig:tsne_classifiers} makes this concrete: under all three fine-tuning
techniques, SF and LV form well-separated clusters, while US and clean exhibit the greatest residual overlap between parameter budgets. The LoRA projection shows much better separation compared to the linear probe, yet the US/clean boundary remains challenging.

\begin{table*}[t]
\centering
  \vspace{0.2em}
\caption{
  Classification results for LV, SF, US, clean; macro-averaged metrics, reported
  as the mean over two seeds.
  GPT-5-mini and Claude Sonnet 4 are evaluated zero-shot and few-shot (4 examples) as baselines.
}
\label{tab:classifier_results}
\begin{tabular}{l|c|ccccc}
\toprule
\textbf{Classifier} & \textbf{Trainable \%} & \textbf{Accuracy} & \textbf{Precision} & \textbf{Recall} & \textbf{F1} & \textbf{MCC} \\
\midrule
GPT-5-mini \small{(zero-shot)}          & 0\%      & 0.46 & 0.490 & 0.460 & 0.469 & 0.281 \\
Claude Sonnet 4 \small{(zero-shot)}     & 0\%      & 0.51 & 0.550 & 0.520 & 0.518 & 0.359 \\
\midrule
GPT-5-mini \small{(few-shot)}           & 0\%      & 0.45 & 0.520 & 0.450 & 0.462 & 0.276 \\
Claude Sonnet 4 \small{(few-shot)}      & 0\%      & 0.55 & 0.620 & 0.560 & 0.557 & 0.412 \\
\midrule
Linear Probe                            & 0.0004\% & 0.75 & 0.750 & 0.745 & 0.748 & 0.665 \\
LoRA Fine-tune \small{(r=16)}           & 0.14\%   & \textbf{0.81} & \textbf{0.810} & \textbf{0.805} & \textbf{0.804} & \textbf{0.745} \\
Full Fine-tune                          & 100\%    & 0.79 & 0.800 & 0.785 & 0.788 & 0.722 \\
\bottomrule
\end{tabular}
\end{table*}

\begin{mdframed}[style=takeawaystyle]
 \textbf{RQ2 Takeaway.}
SpecValidator with LoRA fine-tuning of a 1.5B code model achieves F1=0.804 and MCC=0.745, outperforming both zero-shot and few-shot GPT-5-mini and Claude Sonnet 4 by a wide margin, and also outperforming full fine-tuning. The dominant confusion is between the clean and US classes.
\end{mdframed}

\subsection{Generalization to real (unseen) task descriptions}

The previous RQ results evaluate the classification on test samples that include the mutated data. A natural concern is whether the classifier generalizes beyond this setting to detect real specification issues in the original descriptions. To investigate this, we analyze running the classification on 456 clean descriptions from the three benchmarks, we use the test data of RQ2 (40 HumanEval, 197 MBPP, 219 LiveCodeBench). For each description, the classifier predicts either clean or one of the three defect classes (LV, US, SF); we treat any non-clean prediction as a flagged defect. We then assess whether flagged defects are truly problematic by measuring Pass@1 on the original descriptions across all ten models and by performing a manual inspection.

\begin{figure}[t]
  \centering
  \includegraphics[width=0.98\columnwidth]{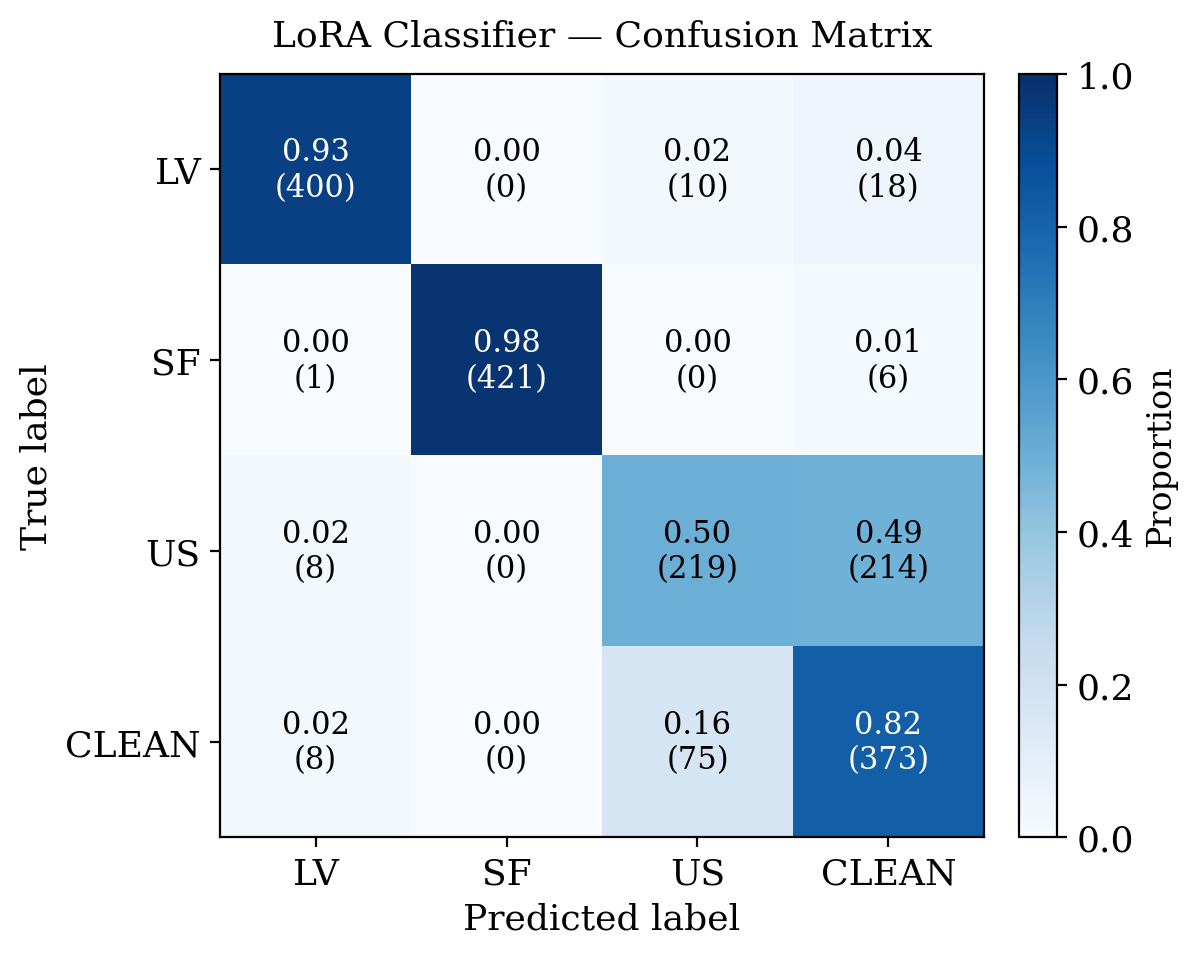}
  \caption{Confusion matrix of the LoRA classifier.}
  \vspace{-0.3em}
  \label{fig:lora_confusion}
\end{figure}

As shown in Figure\ref{fig:lora_confusion}, 83 of the clean descriptions are classified as US (75) and LV(8).  Table~\ref{tab:clean_predictions} reports the classification of clean descriptions from the test set. Of the 456 descriptions, SpecValidator flags 83 (18.2\%) as defective, all of which are Under-Specified: 0 from HumanEval, 33 from MBPP, and 50 from LiveCodeBench. The absence of false positives in HumanEval is consistent with its structured docstring format, which tends to be more explicit than the short MBPP descriptions or the challenging LCB problem statements.

\textbf{The defects flagged by the classifier yield near-zero Pass@1 across all ten models, suggesting that they are originally defective rather than arbitrarily misclassified.}
Table~\ref{tab:classifier_clean_combined} reports Pass@1 on the 83 flagged defective descriptions. For MBPP, 90.9\% of flagged defects fail across all models (failing on an average of 6.7 out of 10); for LiveCodeBench, 98.0\% fail (failing on 7.7 out of 10). These descriptions consistently prevent correct code generation despite being labeled clean in the ground truth benchmark.

\textbf{72.7\% of flagged MBPP descriptions are confirmed as under-specified upon manual inspection.} Three researchers independently reviewed the 33 flagged
MBPP descriptions, which are short enough for human assessment, and 24 of 33 (72.7\%) cases are confirmed as under-specified. Representative examples are shown in Table \ref{tab:clean_as_us_mbpp}.  The 50 LiveCodeBench cases could not be inspected due to the length and complexity of competitive programming statements, but their near-zero Pass@1 is consistent with the same pattern. Taken together, these results indicate that the classifier generalizes beyond the injected defects and indicates latent quality issues in the benchmark ground truth.

\begin{mdframed}[style=takeawaystyle]
\textbf{RQ3 Takeaway.} SpecValidator flags 83 of 456 clean benchmark descriptions as under-specified. These flagged descriptions yield extremely low Pass@1 across all ten models (90.9\% failure rate on MBPP, 98.0\% on LiveCodeBench), and 72.7\% of inspectable MBPP cases are confirmed as under-specified upon manual review. These results indicate that SpecValidator generalises beyond its training distribution and detects specification defects in benchmark ground truth.
\end{mdframed}

\section{Threats to Validity}
 
\subsection*{Internal validity}
 
A primary validity concern is whether our mutants are genuinely defective rather than arbitrarily corrupted. We mitigate this through a two-tier validation strategy: An independent LLM judge (Qwen2.5-Coder-32B) scores all 6{,}573 mutants for compliance and naturalness, and a stratified manual sample of 100 instances is manually checked. Human annotators closely rate the automated scores, providing confidence on the representativeness of our mutations. Another concern is the use of greedy decoding (\texttt{temperature\,=\,0}) during inference. While this ensures reproducibility, it may underestimate the variance introduced by temperature sampling and overestimate sensitivity for problems where the model's top-1 beam is near the decision boundary. We treat this as a conservative choice: if greedy decoding shows sensitivity, sampling would only amplify it.
 
\subsection*{External validity}
 
The finding that LCB is immune to prompt-level perturbations may not generalize well. Conversely, findings on HumanEval and MBPP may not transfer to more complex, library-intensive tasks such as those in BigCodeBench~\cite{zhuo2025bigcodebench}, where the relationship between prompt precision and functional correctness is likely to differ. All open-source models are evaluated in their instruction-tuned variants; base models may exhibit a different and potentially higher sensitivity. 
Our findings on model-family-specific behavior (e.g., the unusually low LV sensitivity of Qwen2.5-Coder-32B) should be interpreted with caution as it may be architecture- or training-data specific. Furthermore, although we include HumanEval and MBPP despite their known contamination risks~\cite{matton2024leakage,
riddell2024contamination}, contamination may artificially compress the measured sensitivity: if models have partially memorized solutions, they may be more robust to prompt changes than they would otherwise be. Including LCB as a contamination-free benchmark is intended to mitigate this threat. 
 
\subsection*{Construct validity}
 
We use Pass@1 as a correctness metric, which treats all failures identically. Pass@1 is the standard in the code generation literature and reflects the real-world scenario in which a developer accepts the model's first suggestion, but it cannot distinguish near misses from total failures. 
User prompts may contain mixed or compound defects that do not fall cleanly into our categories. The judgment of one constraint removed for US mutations is necessarily subjective; despite our quality-control pipeline, some US mutants in MBPP were rated as under-specified even in their original (CLEAN) form, a limitation of our LoRA classifier's false-positive analysis explicitly surfaces as a finding rather than an error. 

\begin{table}[t]
\caption{SpecValidator on the 456 real unseen description. TN = correctly predicted CLEAN; FP = incorrectly flagged as defective (all as US). Specificity = TN\,/\,Total; FP rate = FP\,/\,Total.}
\vspace{-0.2em}
\label{tab:clean_predictions}
\centering
\begin{tabular}{lrrrrrrr}
\toprule
Benchmark & Total & TN & FP & Specificity & FP Rate \\
\midrule
HumanEval     & 40  & 40  & 0  & 100.0\% & 0.0\%  \\
MBPP          & 197 & 164 & 33 & 83.2\%  & 16.8\% \\
LiveCodeBench & 219 & 169 & 50 & 77.2\%  & 22.8\% \\
\midrule
Total         & 456 & 373 & 83 & 81.8\%  & 18.2\% \\
\bottomrule
\end{tabular}
\vspace{-0.2em}
\end{table}

\begin{table}[t]
\centering
\small
\vspace{-0.2em}
\caption{Pass@1 results on CLEAN test samples predicted as non-CLEAN by the classifier.
         \textbf{F} = pass@1 False (failing); \textbf{P} = pass@1 True (passing).
         The \emph{Misclassified} row aggregates across all samples;
         the parenthetical reports the average number of models on which
         a misclassified sample fails.}
\label{tab:classifier_clean_combined}
\begin{tabular}{lrrrr}
\toprule
\textbf{Model} & \textbf{N} & \textbf{F} & \textbf{P} & \textbf{\%F} \\
\midrule
\multicolumn{5}{l}{\textit{MBPP}} \\
\midrule
CodeLlama-34B       & 33 & 25 &  8 & 75.8 \\
CodeLlama-7B        & 33 & 24 &  9 & 72.7 \\
StarCoder2-15B      & 33 & 23 & 10 & 69.7 \\
DeepSeek-Coder-33B  & 33 & 21 & 12 & 63.6 \\
Qwen2.5-Coder-32B   & 33 & 20 & 13 & 60.6 \\
DeepSeek-Coder-6.7B & 33 & 20 & 13 & 60.6 \\
Codestral-22B       & 33 & 20 & 13 & 60.6 \\
GPT-5-mini         & 33 & 19 & 14 & 57.6 \\
Qwen2.5-Coder-7B    & 33 & 15 & 18 & 45.5 \\
Claude Sonnet       & 33 & 15 & 18 & 45.5 \\
\cmidrule(lr){1-5}
\textit{Misclassified} (Fail on avg 6.7/10 models) & 33 & 30 & 3 & 90.9 \\
\midrule
\multicolumn{5}{l}{\textit{LiveCodeBench}} \\
\midrule
StarCoder2-15B      & 50 & 49 &  1 & 98.0 \\
CodeLlama-7B        & 50 & 45 &  5 & 90.0 \\
CodeLlama-34B       & 50 & 43 &  7 & 86.0 \\
DeepSeek-Coder-33B  & 50 & 43 &  7 & 86.0 \\
DeepSeek-Coder-6.7B & 50 & 41 &  9 & 82.0 \\
Codestral-22B       & 50 & 41 &  9 & 82.0 \\
Qwen2.5-Coder-7B    & 50 & 39 & 11 & 78.0 \\
Qwen2.5-Coder-32B   & 50 & 36 & 14 & 72.0 \\
GPT-5-mini         & 50 & 27 & 23 & 54.0 \\
Claude Sonnet       & 50 & 15 & 35 & 30.0 \\
\cmidrule(lr){1-5}
\textit{Misclassified} (Fail on avg 7.7/10 models) & 50 & 49 &  1 & 98.0 \\
\bottomrule
\end{tabular}
\vspace{-0.2em}
\end{table}

\begin{table}[t]
\centering
\small
\caption{Example \textsc{Under-Specified} (real) defects detected by SpecValidator.}
\vspace{-0.2em}
\label{tab:clean_as_us_mbpp}
\begin{tabular}{p{0.6cm} p{3cm} p{3cm}}
\toprule
\textbf{ID} & \textbf{Prompt} & \textbf{Ambiguity} \\
\midrule
MBPP 277 & Filter a dictionary based on values. 
         & By what condition? A threshold? \\
\addlinespace
MBPP 466 & Find the peak element in the given array. 
         & Local peak or global max? Index or value? \\
\addlinespace
MBPP 382 & Find the number of rotations in a circularly sorted array. 
         & Left or right rotation? Ascending or descending? \\

\bottomrule
\end{tabular}
\vspace{-0.2em}
\end{table}

\section{Conclusion}

This study investigates the impact of task description defects on the LLM-based code generation and ways to automatically detect such defects. We find that defect impact is not uniform: it depends jointly on defect type and benchmark specification structure. Under-Specification is the most damaging class, causing Pass@1 drops of up to 15.3\% with no model immune, including state-of-the-art reasoning models above 95\% baseline. Lexical Vagueness produces moderate, benchmark-dependent degradation, whereas surface formatting noise has a negligible effect on code-specialized models. Benchmarks with richer contextual grounding, such as LiveCodeBench, are substantially more resilient to prose-level defects, establishing the specification structure as a first-order moderator of robustness that has been invisible in prior work.

On the detection side, SpecValidator, a lightweight LoRA-fine-tuned 1.5B classifier, achieves F1\,=\,0.804 and MCC\,=\,0.745, outperforming few-shot GPT 5-mini and Claude Sonnet~4 by a wide margin despite updating only 0.14\% of parameters, confirming that strong code generation ability does not transfer to the prompt meta-reasoning. More importantly, SpecValidator generalises beyond its training distribution: it flags 83 clean benchmark descriptions as under-specified, 72.7\% of which are confirmed as genuinely defective upon manual inspection, with near-zero Pass@1 across all ten models supporting the same conclusion for the remaining cases. This demonstrates that SpecValidator can surface latent quality issues in benchmark ground truth, a finding with direct implications for how code generation benchmarks are constructed and evaluated.

Taken together, these results argue that description defects should be treated as structurally distinct phenomena requiring targeted fixing. Robustness evaluations that aggregate across perturbation types or rely exclusively on simple single-source benchmarks risk drawing conclusions that do not transfer to realistic settings. We release our dataset of 6,573 defective task descriptions to support reproducibility and future work.




\clearpage
\bibliographystyle{ACM-Reference-Format}
\bibliography{base.bib}

\appendix

\end{document}